\documentclass[11pt]{revtex4-1}
\usepackage{natbib}
\usepackage{amssymb,amsmath,xcolor,graphicx,xspace,colortbl,textcomp, rotating}

\usepackage{color}

\begin{document}

\title{Energy in Higher-Derivative Gravity via Topological Regularization}
\author{Gaston Giribet\smallskip}
\affiliation{Center for Cosmology and Particle Physics,\\ New York University, 726 Broadway, 10003 New York City, USA.}
\email{gg1043@nyu.edu}
\author{Olivera Miskovic \smallskip}
\affiliation{Instituto de F\'{\i}sica,\\ Pontificia Universidad Cat\'olica de Valpara\'{\i}so, Casilla 4059, Valparaiso, Chile.}
\email{olivera.miskovic@pucv.cl}
\author{Rodrigo Olea}
\affiliation{Departamento de Ciencias F\'{\i}sicas,\\ Universidad Andres Bello, Sazi\'e 2212,\\ Piso 7, Santiago, Chile.}
\email{rodrigo.olea@unab.cl, \; d.riverabe@gmail.com}
\author{David Rivera-Betancour\smallskip}
\affiliation{Departamento de Ciencias F\'{\i}sicas,\\ Universidad Andres Bello, Sazi\'e 2212,\\ Piso 7, Santiago, Chile.}
%\email{}

\begin{abstract}
\vspace{1cm}

We give a novel definition of gravitational energy for an arbitrary theory of gravity including quadratic-curvature corrections to Einstein equations. We focus on the theory in four dimensions, in presence of negative cosmological constant, and with asymptotically anti-de Sitter (AdS) boundary conditions. As a first example, we compute the gravitational energy and angular momentum of Schwarzschild-AdS black holes, for which we obtain results consistent with previous computations performed using different methods. However, our method differs qualitatively from other ones in the feature of being intrinsically non-linear. It relies on the idea of adding to the gravity action topological invariant terms which suffice to regularize the Noether charges and render the variational problem well-posed. This is an idea that has been previously considered in the case of second-order theories, such as general relativity and which, as shown here, extends to higher-derivative theories. Besides black holes, we consider other solutions, such as gravitational waves in AdS, for which we also find results in agreement. This enables us to investigate the consistency of this approach in the non-Einstein sector of the theory.
\end{abstract}

\maketitle

\section{Introduction}

Topology plays a fundamental role in theoretical physics, from the theory of fundamental particles to the study of topological phases of matter. It is a central notion in the study of dynamical systems and, from there, it extends to practically every area of physics. Derived notions, such as topological classes and topological invariants are essential tools in gauge field theories and quantum gravity, being indispensable for the study of anomalies, of quantum transitions, and of topological defects. In the case of the theory of gravity, being a theory of the geometry of the spacetime itself, the role played by topological invariants is particularly important. They appear in relation to the Euclidean path integral formulation of the theory, in the study of gravitational instantons, and in higher-dimensional, higher-curvature extensions of the theory. Here, following an idea previously explored by one of the authors and collaborators in the case of second-order theories such as general relativity \cite{ACOTZ}, we propose a novel use of topological invariants in the context of gravity, namely as regulators of the Noether charge computation in theories that include higher-derivative terms. This is a very important problem, as higher-derivative corrections to Einstein theory naturally arise when quantum effects are taken into account, and therefore having a method to compute the observables in those cases is crucial. For theories of this sort, we show that the Euler characteristic, which in virtue of the Chern-Weil-Gauss-Bonnet theorem is expressed in terms of the Riemann tensor, can be used to regularize the infrared divergences that typically appear when trying to compute the conserved Noether charges in asymptotically, locally maximally symmetric spaces. Such divergences {arise} due to the non-compactness of the spacetime but, as we will show, the inclusion of the topological invariant in the action suffices to render the result finite and the variational problem well-posed. We refer to this phenomenon as topological regularization. As said, this has been observed to occur in theories with second-order field equations, such as general relativity (GR). Here, we show that this is, indeed, a much more general phenomenon, which also occurs in higher-derivative theories with arbitrary quadratic corrections to the Einstein-Hilbert action. Having such a non-linear method to compute charges in theories with actions that are quadratic in the curvature tensor and in presence of cosmological constant is very important, as such theories are known to suffer from linear instabilities that make the linear analysis to break down.

%\bc{[\textbf{We propose to eliminate this paragraph for so short paper.}] This paper is organized as follows: In section II, we review some properties of higher-curvature theories that are necessary for our analysis. In section III, we discuss different definitions of gravitational energy for {quadratic-}curvature gravity theories that have been proposed in the literature. We review the so-called Abbott-Deser-Tekin formalism as well as the Iyer-Wald formalism to compute Noether charges. In section IV, we consider the inclusion of a topological invariant term as a regulator to compute the Noether charges and render the variational problem well-posed. This will provide us with an intrinsically non-linear method to compute the gravitational energy and angular momentum of black hole solutions. Also in section IV, we focus on gravitational waves that correspond to non-Einstein solutions of the theory. As for the case of black holes, the computation of gravitational energy for gravitational waves solutions yields a result in agreement with other methods. This enables us to probe our method in the non-Einstein sector of the theory. We also discuss the case of asymptotically Lifshitz black holes. Section V contains our conclusions.}

\section{Quadratic-curvature gravity}

Quantum effects induce higher-derivative corrections to the effective gravitational action. This is particularly realized in string theory, where such corrections can be computed explicitly to different orders. In the case of heterotic string theory, for example, $R^2$ corrections are seen to appear at the next-to-leading order in the $\alpha ' $-expansion of the low energy effective action. $R^2$ terms also arise in different string theory compactifications. Here, aiming at not restricting ourselves to any specific model, we will consider the most general {quadratic-}curvature gravity (QCG) theory in four dimensions and in presence of cosmological constant. This is described by an action $I=(16\pi G)^{-1}\int d^4x\sqrt{-g}\mathcal{L}$ with a Lagrangian density given by an arbitrary combination of Ricci-squared and
scalar curvature-squared terms, namely
\begin{equation}
\mathcal{L=}R-2\Lambda +\alpha R^{\mu \nu }R_{\mu
\nu }+\beta R^{2} \,.  \label{quadratic}
\end{equation}

The cosmological constant $\Lambda$, which here we consider negative, relates to the radius $\ell$ of the anti-de Sitter (AdS) solution by $\Lambda =-3/\ell ^{2}$. In (\ref{quadratic}), $\alpha$ and $\beta$ are two arbitrary coupling constants of mass dimension $-2$, which set the length scale $l=\max \{\sqrt{| \alpha |}, \sqrt{| \beta |}\}$ at which the higher-curvature effects start to dominate over the Einstein-Hilbert term. Action (\ref{quadratic}) does not include the Kretschmann scalar (Riemann-squared term), as in four dimensions its presence can always be compensated by a redefinition of the couplings $\alpha $ and $\beta $, up to a boundary term.

As already mentioned, non-linear terms in the Riemann tensor are expected to {appear} in the quantum gravity effective action. The presence of such terms improves the ultraviolet behavior of the theory relative to GR. In fact, the theory defined by Lagrangian (\ref{quadratic}) turns out to be renormalizable \cite{Stelle}. The price to be paid for this is the emergence of ghost-like additional degrees of freedom. The class of theories given by the Lagrangian (\ref{quadratic}) leads to
fourth-order field equations which generically describe a massive scalar mode and a massive spin-2
field, in addition to the massless graviton already present in GR.

Taking arbitrary variations of the action yields the equation of
motion (EOM) of the theory plus a surface term,
\begin{equation}
\delta I=\frac{1}{16\pi G}\int\limits_{M}d^{4}x\sqrt{-g}\left( E^{\mu \nu
}\delta g_{\mu \nu }+\nabla _{\alpha }\Theta ^{\alpha }\right) \, ,
\label{variation}
\end{equation}%
where the field equations are given by $E^{\mu \nu }=0$, with the symmetric rank-2 tensor $E^{\mu \nu }=G^{\mu \nu }+\alpha E_{(\alpha )}^{\mu \nu }+\beta E_{(\beta )}^{\mu \nu }$ being composed by the Einstein tensor
\begin{equation}
G^{\mu \nu }=R^{\mu \nu }-\frac{1}{2}\,Rg^{\mu \nu }+\Lambda g^{\mu \nu }\,,
\label{Einstein}
\end{equation}%
supplemented with the two contributions coming from the quadratic-curvature terms,
\begin{equation}
E_{(\alpha )}^{\mu \nu }=\left( g^{\mu \nu }\Box -\nabla ^{\mu }\nabla ^{\nu
}\right) R+\Box \left( R^{\mu \nu }-\frac{1}{2}\,g^{\mu \nu }R\right) +2\left(
R^{\mu \sigma \nu \rho }-\frac{1}{4}\,g^{\mu \nu }R^{\sigma \rho }\right)
R_{\sigma \rho }\,,  \label{E1}
\end{equation}%
and
\begin{equation}
E_{(\beta )}^{\mu \nu }=2R\left( R^{\mu \nu }-\frac{1}{4}\,g^{\mu \nu
}R\right) +2\left( g^{\mu \nu }\Box -\nabla ^{\mu }\nabla ^{\nu }\right) R.
\label{E2}
\end{equation}

For a generic gravity theory, surface terms can be found
as partial derivatives of the Lagrangian with respect to the Riemann
tensor in the form of
\begin{equation}
\Theta ^{\alpha }(\delta g,\delta \Gamma )=2E_{\mu \nu }^{\alpha \beta
}\,g^{\nu \lambda }\delta \Gamma _{\beta \lambda }^{\mu }+2\nabla ^{\mu
}E_{\mu \nu }^{\alpha \beta }\left( g^{-1}\delta g\right) _{\beta }^{\nu }\,,
\label{surface}
\end{equation}%
where the rank-4 tensor $E_{\mu \nu }^{\alpha \beta }$ is defined by
\begin{equation}
E_{\mu \nu }^{\alpha \beta }=\frac{\partial \mathcal{L}}{\partial R_{\alpha
\beta }^{\mu \nu }}\,.  \label{derivative}
\end{equation}
For Lagrangians (\ref{quadratic}), this functional derivative takes the form
\begin{equation}
E_{\mu \nu }^{\alpha \beta }={\frac{1}{2}}\left( \delta _{\lbrack \mu
\nu ]}^{[\alpha \beta ]}+\alpha R_{[\mu }^{[\alpha }\delta _{\nu ]}^{\beta
]}+2\beta R\delta _{\lbrack \mu \nu ]}^{[\alpha \beta ]}\right) \ ,
\label{Etensor}
\end{equation}
where $\delta _{\lbrack \mu \nu ]}^{[\alpha \beta ]}$ stands for the totally anti-symmetric product of Kronecker tensors {with normalization factor equal to unity.}

The vacuum states of the theory correspond to maximally symmetric spaces
\begin{equation}
R_{\mu \nu }^{\alpha \beta }=-\frac{1}{\ell^{2}}\,\delta
_{\lbrack \mu \nu ]}^{[\alpha \beta ]} \ , \label{globalAdS}
\end{equation}
where $\ell $ is the curvature radius of the space. In four dimensions, the addition of quadratic terms does not alter the relation between $\Lambda $ and $\ell$, which remains $\Lambda=-3/\ell^2$, independent of $\alpha$ and $\beta$. Besides, four dimensions is also special in the sense that all Einstein spacetimes (i.e. $R_{\mu \nu }=-3/\ell ^{2}g_{\mu \nu }$) persist as solutions of theory. Notice that for QCG, the second contribution on the right hand side of Eq.(\ref{surface}) vanishes identically for Einstein spaces.

The particle content of the theory can be obtained from the linearization of
the field equations (\ref{Einstein})-(\ref{E2}) around the AdS
background metric $\bar{g}_{\mu \nu }$. By writing $g_{\mu \nu }=\bar{g}_{\mu \nu
}+h_{\mu \nu }$, where $h_{\mu \nu }$ is a small
perturbation \citep{Deser:Tekin}, the linearized equations of motion read
\begin{eqnarray}
\delta \left( G_{\mu \nu }+E_{\mu \nu }\right) &=&\left[ 1+2\Lambda \left(
\alpha +4\beta \right) \right] G_{\mu \nu }^{L}+\alpha \left[ \left( \bar{%
\square}-\frac{2\Lambda }{3}\right) G_{\mu \nu }^{L}-\frac{2\Lambda }{3}\,R^{L}%
\bar{g}_{\mu \nu }\right] +  \notag \\
&&+\left( \alpha +2\beta \right) \left( -\bar{\nabla}_{\mu }\bar{\nabla}%
_{\nu }+\bar{g}_{\mu \nu }\bar{\square}+\Lambda \bar{g}_{\mu \nu }\right)
R^{L}\,,  \label{EOMLin}
\end{eqnarray}%
where $G_{\mu \nu }^{L}$ and $R^{L}$ mean the linearized version of
the Einstein tensor and Ricci scalar, respectively. The bar denotes covariant derivatives with
respect to the background metric, $\bar{g}_{\mu \nu }$.

In order to simplify the analysis of the linearized equations, it is convenient to choose the transverse
traceless gauge $\bar{\nabla}_{\mu }h^{\mu \nu }=0$. Then, from
the trace of (\ref{EOMLin}) one gets the equation for the propagation of the
scalar mode, namely
\begin{equation}
h-2\left( \alpha +3\beta \right) \bar{\Box}h =0\,,
\label{scalar_mode}
\end{equation}%
where $h=\bar{g}^{\mu \nu }h_{\mu \nu }$ is the trace of the perturbation, which describes the scalar particle of the theory. For the special choice $\alpha=-3\beta$, this degree of freedom becomes strongly coupled and disappears from the spectrum.

\section{Energy definition in QCG}

\subsection{Linearized perturbations and charges}

A definition of gravitational energy for generic QCG theories was given by Deser and Tekin in Refs.\cite{Deser:Tekin,Deser:BTekin} by extending the work of Abbott and Deser, who had achieved in \cite{Abbott:Deser} such an energy definition for Einstein theory in presence of cosmological constant. This method is thus often referred to as Abbott-Deser-Tekin formalism or simply as ADT method. The procedure is based on the
linearization of the equations of motion (EOM) around a globally (A)dS (background) metric. The method is also valid in the case of vanishing cosmological constant. The linearized charges obtained in this way are expressed as the integral over an orientable codimension-2 surface $\Sigma $,
\begin{equation}
Q_{_{\rm{ADT}}}^{\alpha }[\bar{\xi}]={\frac{1}{8\pi G}}\int\limits_{\Sigma
}dS_{\beta }\,q_{_{\rm{ADT}}}^{\alpha \beta }\,,  \label{ADTcharge}
\end{equation}
where $dS_{\nu }$ is the differential area element defined by the exterior unit vector normal to the surface and where $\Sigma$ defines the integration at fixed time and fixed radial coordinate.
The integrand is given by the components of the prepotential
\begin{eqnarray}
q_{_{\rm{ADT}}}^{\alpha \beta } &=&\left[ 1+2\Lambda \left( \alpha +4\beta
\right) \right] \left( \bar{\xi}_{\lambda }\bar{\nabla}^{[\alpha }h^{\beta
]\lambda }+\bar{\xi}^{[\alpha }\bar{\nabla}^{\beta ]}h+h^{\lambda \lbrack
\alpha }\bar{\nabla}^{\beta ]}\bar{\xi}_{\lambda }-\bar{\xi}^{[\alpha }\bar{%
\nabla}_{\lambda }h^{\beta ]\lambda }+\frac{1}{2}h\bar{\nabla}^{\alpha }\bar{%
\xi}^{\beta }\right)  \notag \\
&&+\left( \alpha +2\beta \right) \left( 2\bar{\xi}^{\,[\alpha }\bar{\nabla}%
^{\beta ]}R^{L}+R^{L}\bar{\nabla}^{\alpha }\bar{\xi}^{\beta }\right) -\alpha
\left( 2\bar{\xi}_{\lambda }\bar{\nabla}^{[\alpha }G_{L}^{\beta ]\lambda
}+2G_{L}^{\lambda \lbrack \alpha }\bar{\nabla}^{\beta ]}\bar{\xi}_{\lambda
}\right) \ ,  \label{ADTF}
\end{eqnarray}
where $\bar{\xi}_{\lambda }$ are the components of the Killing vector that generates the isometry associated to the Noether charge that is being computed. As before, the subscript ${L}$ stands for the linearized version of every curvature tensor involved.

The ADT method has shown to be useful to compute the Noether charges associated to several solutions of physical importance. In the case of Schwarzschild-AdS black hole, for example, whose metric in standard Schwarzschild coordinates is given by
\begin{equation}
ds^2=-\Big(1-\frac{2mG}{r}+\frac{r^2}{\ell^2}\Big)dt^2 + \Big(1-\frac{2mG}{r}+\frac{r^2}{\ell^2}\Big)^{-1}dr +r^2 (\sin^2\theta \ d\phi^2 + d\theta^2)
\end{equation}
with $t\in \mathbb{R}$, $r\in \mathbb{R}_{\geq 0}$, $\phi\in [0,2\pi]$, $\theta\in [0,\pi]$, the ADT gravitational energy computation yields
\begin{equation}
M={Q_{_{\rm{ADT}}}^0[\partial_t]}=m\left[ 1+2\Lambda (\alpha +4\beta )\right] ,  \label{Mass}
\end{equation}%
which, in particular, tends to the usual mass parameter $m$ in the absence of higher-curvature terms ($\alpha=\beta=0$) or in absence of cosmological constant ($\Lambda=0$). In general, there is no need to go beyond linear order in the curvature, as
the energy is linear in $m$. Thus, even if the Lagrangian contains higher
powers in the curvature, the conserved charge can be consistently truncated
to an expression linear in $R$ if the corresponding background is not
degenerate. An exception to this happens at the so-called \textit{critical points}, which are special points of the space of coupling constants where the massive spin-2 field actually becomes massless and the spin-0 mode becomes strongly coupled. At such a critical point, the black hole solutions of the QCG theory have vanishing energy and vanishing angular momentum, and the fourth order linearized equations become confluent and develop low-decaying extra modes. This is, for instance, the case of Critical Gravity \cite{Lu:Pope}, which is defined by choosing in (\ref{quadratic}) the values $\alpha =-3\beta =3/(2\Lambda )$. We have already mentioned that when $\alpha =-3\beta $ the scalar mode in (\ref{scalar_mode}) is eliminated and $h=0$ becomes the only possible solution. Then, Eq.(\ref{EOMLin}) reduces to
\begin{equation}
\left( \bar{\Box}-\frac{2\Lambda }{3}\right) \left( \bar{\Box}-\frac{%
2\Lambda }{3}-\frac{2\Lambda \beta +1}{3\beta }\right) h_{\mu
\nu }=0\,,  \label{spin2}
\end{equation}
which describes the propagation of a massless graviton and a massive spin-2 mode. If, in addition, one imposes the particular value $\beta =-1/(2\Lambda )$, then the graviton degenerates with the additional spin-2 mode. Under these conditions, the formula (\ref{Mass}) for the mass of the AdS-Schwarzschild black holes yields a vanishing result \cite{Lu:Pope}. This result, however, has to be confirmed by going to next-to-leading order in perturbation theory as, because of the degeneracy in the spin-2 fields, QCG theory at the critical point may present linear instabilities \cite{Tekin:Altas} and thus the linear analysis can not be fully trusted. This is, indeed, one of the main motivations we have to look for an intrinsically non-linear method to compute conserved charges in QCG, what we will do in the following section.

\subsection{Diffeomorphic charges}

A preliminary ingredient for our discussion is the analysis of the Noether currents carried out by Wald and collaborators for the theory of gravity. Assuming the Lagrangian to be a function of the metric and the curvature
tensor, $\mathcal{L}(g_{\mu \nu },R_{\mu \nu \alpha \beta })$, the
corresponding Noether current adopts the form
\begin{equation}
J^{\alpha }=2E_{\mu \nu }^{\alpha \beta }\,g^{\nu \lambda }\delta
_{\xi }\Gamma _{\beta \lambda }^{\mu }+2\nabla ^{\mu }E_{\mu \nu }^{\alpha
\beta }\left( g^{\nu \lambda }\delta _{\xi }g_{\lambda \beta }\right) +\mathcal{L}\,\xi ^{\alpha }\,,
\label{current]}
\end{equation}%
where $\delta _{\xi }$ stands for the Lie derivative acting on the fields,
\begin{eqnarray}
\mathcal{\delta }_{\xi }\,\Gamma _{\beta \lambda }^{\mu } &=&-\frac{1}{2}%
\left( \nabla _{\beta }\nabla _{\lambda }\xi ^{\mu }+\nabla _{\lambda
}\nabla _{\beta }\xi ^{\mu }\right) - R_{\ \lambda \sigma \beta
}^{\mu } \xi ^{\sigma } \,,   \nonumber \\
\delta _{\xi }\,g_{\lambda \beta } &=&-\left(\nabla _{\lambda }\xi _{\beta }+\nabla
_{\beta }\xi _{\lambda }\right)\,.  \label{Lieg}
\end{eqnarray}
In contrast to the original derivation in Refs.\cite{Iyer:Wald,Wald:Zoupas}, here we will not take the Killing vector condition
for the diffeomorphism $\xi ^{\mu }$ at this step. {This allows us to account for contributions coming from derivatives of the tensor $E_{\mu \nu }$.} After some algebraic
manipulation, the Noether current is written down as a total derivative plus
the field equations contracted with the vector $\xi $ \cite{Maedaetal},
\begin{equation}
J^{\alpha }=2\nabla _{\beta }\left( E_{\mu \nu }^{\alpha \beta }\nabla ^{\mu
}\xi ^{\nu }+2\nabla ^{\mu }E_{\mu \nu }^{\alpha \beta }\xi ^{\nu }\right)
\,+(EOM)^{\alpha \beta } \xi_{\beta } \,.  \label{NWcurrent}
\end{equation}%
The above equation implies that the conserved charge associated to an
isometry generated by a Killing vector $\xi^{\mu}$ is an integral over a codimension-2 surface $\Sigma $,
\begin{equation}
Q_{_{\rm{IW}}}^{\alpha }[\xi ]={\frac{1}{8\pi G}}\int\limits_{\Sigma }dS_{\beta }\,q_{_{\rm{IW}}}^{\alpha \beta} \,,  \label{NWcharge}
\end{equation}
for the Iyer-Wald prepotential
\begin{equation}
q_{_{\rm{IW}}}^{\alpha \beta}=
{E_{\mu \nu }^{\alpha \beta }\nabla ^{\mu }\xi ^{\nu }+2\nabla ^{\mu }E_{\mu
\nu }^{\alpha \beta }\xi ^{\nu }}\,.
\label{NWcharge2}
\end{equation}

{The above formula gives rise to the standard black hole entropy when evaluated at the horizon. However, when it comes to
the discussion of charges in the asymptotic region, it is  necessary to correct the prepotential as $q_{_{\rm{IW}}}^{\alpha \beta}\rightarrow q_{_{\rm{IW}}}^{\alpha \beta}+\xi^{[\alpha}B^{\beta]}$, where $B$ is a boundary term whose form can be found assuming suitable boundary conditions.
It is only after the contribution of such term is added that this construction reproduces the results as the canonical formalism \cite{Iyer:Wald}.}

There exist other methods to compute Noether charges suitable to be adapted to the case of Lagrangians with higher-derivative couplings. One particularly strong method is the so-called Barnich-Brandt method \cite{BarnichBrandt}, which has the advantage of being fully constructive. This method is often used to compute charges associated to asymptotic isometries, including possible central terms in the charge algebra. In the particular case of exact (in opposition to asymptotic) Killing vectors, the method of \cite{BarnichBrandt} agrees with the Iyer-Wald computation, and in the case of black holes it reproduces the ADT charges.

{In the following section, we propose a qualitatively different way of regularizing the Iyer-Wald charges for the Lagrangian (\ref{quadratic}). This amounts to trade off the effect of $B$ by a correction in $q_{_{\rm{IW}}}^{\alpha \beta}$ due to a topological term in the action.}

\section{Topological term and conserved charges}

Commonly omitted from the gravity action {in four dimensions}, the Pfaffian
\begin{equation}
\mathcal{E}_{4}=\sqrt{-g}\left( R_{\alpha \beta }^{\mu \nu }R_{\mu \nu
}^{\alpha \beta }-4R^{\mu \nu }R_{\mu \nu }+R^{2}\right) \,,
\label{GaussBonnet}
\end{equation}
(hereafter referred to as the Gauss-Bonnet term) can be suitably added to the Lagrangian (\ref{quadratic}) with a dimensionful  coupling $\gamma $. This is, indeed, a topological term and, as such, it does not affect the dynamics. However, its presence does modify the infrared divergences and the Noether currents. In other words, with the addition of $\gamma \mathcal{E}_{4}$ to {$ \sqrt{-g} {\mathcal L}$,} the bulk field equations $E^{\mu\nu}=0$ do not change, but the
surface term does. The latter appears augmented by the term
\begin{equation}
\Theta _{\rm{GB}}^{\alpha }={\gamma \,} g^{\nu \lambda }\delta
\Gamma _{\beta \lambda }^{\mu }\delta _{\lbrack \mu \nu \sigma \lambda
]}^{[\alpha \beta \gamma \delta ]}R_{\gamma \delta }^{\sigma \lambda }\,,
\label{DeltaTheta}
\end{equation}%
where, again, we are using the totally-antisymmetric Kronecker delta symbol as a
shorthand. The minimum requirement for (\ref{DeltaTheta}) is the total action to attain a well-posed variational principle when evaluated on a vacuum state of the theory. That is, one demands $\delta I_{\rm{total}}=0$ on the maximally symmetric solution, namely global AdS$_4$. The total surface term $\Theta _{\rm{total}}^{\alpha }=\Theta_{\rm{QCG}}^{\alpha }+\Theta _{\rm{GB}}^{\alpha }$ on AdS$_4$ takes the form
\begin{equation}
\Theta _{\rm{total}}^{\alpha }=\delta _{[ \mu \nu
]}^{[\alpha \beta ]}\,g^{\nu \lambda }\delta \Gamma _{\beta \lambda }^{\mu
}\left[ 1+2\Lambda \left( 4\beta +\alpha \right) -\frac{4\gamma }{\ell ^{2}}%
\right] \ ,  \label{surface_AdS}
\end{equation}%
and demanding this to vanish, amounts to unambiguously {setting} the coupling of the topological term to
\begin{equation}
\gamma =\frac{\ell ^{2}}{4}\left[ 1+2\Lambda \left( 4\beta +\alpha \right)
\right] \,.  \label{coupling}
\end{equation}
This comes to generalize the result of \cite{ACOTZ}, to which it reduces when $\alpha=\beta=0$. In other words, (\ref{coupling}) expresses the fact that the coupling constant $\gamma $ includes the effect of the quadratic terms in renormalizing the Newton constant. In the case $\alpha=\beta=0$, (\ref{coupling}) yields $\gamma=-3/(4\Lambda )$, which is the precise value that, when formulated the theory in 5 dimensions, the coupling constant $\gamma $ has to take for the Einstein-Gauss-Bonnet action to admit to be written as a Chern-Simons gauge theory.

Besides the condition on the vacuum state $\Theta _{\text{total}}^{\alpha }=0$ for the variation of the action to vanish on AdS$_4$, what is not {\it a priori} clear is whether or not such value of $\gamma $ is also the one that produces a finite variation of the action when one departs from the global AdS space and considers, for instance, massive solutions instead. Below, we will show for physically relevant solutions that the addition of the Gauss-Bonnet term (\ref{GaussBonnet}) with (\ref{coupling}) actually suffices to render finite the energy definition coming from Noether
theorem.

Indeed, once the topological term is taken into account, the prepotential (\ref{NWcharge2}) turns into
\begin{eqnarray}
q_{\rm{top}}^{\alpha \beta } &=&\frac{1}{2}\nabla ^{\mu }\xi ^{\nu }\left[ \delta
_{\lbrack \mu \nu ]}^{[\alpha \beta ]}+\alpha R_{[\mu }^{[\alpha }\delta
_{\nu ]}^{\beta ]}+2\beta R\delta _{\lbrack \mu \nu ]}^{[\alpha \beta ]}+%
\frac{\ell ^{2}}{4}\left(\rule{0pt}{12pt} 1+2\Lambda \left( 4\beta +\alpha \right) \right)
\delta _{\lbrack \mu \nu \sigma \lambda ]}^{[\alpha \beta \gamma \delta
]}R_{\gamma \delta }^{\sigma \lambda }\right] +  \notag \\
&&+\nabla ^{\mu }\left( \alpha R_{[\mu }^{[\alpha }\delta _{\nu ]}^{\beta
]}+2\beta R\delta _{\lbrack \mu \nu ]}^{[\alpha \beta ]}\right) \xi ^{\nu
}\,,  \label{prepoDavid}
\end{eqnarray}
and the associated charge (\ref{NWcharge}) for the quadratic-curvature
gravity action is expressed as a surface integral
\begin{equation}
Q_{\rm{top}}^{\alpha }[\xi ]={\frac{1}{8\pi G}}\int\limits_{\Sigma
}dS_{\beta }\,\,q^{\alpha \beta }_{\rm{top}}\,.  \label{RenormalizedCharge}
\end{equation}
The subscript $\text{top}$ refers to the topological origin of the supplementary term (\ref{GaussBonnet}).

\subsection{Black hole gravitational energy}

Restricting ourselves first to the Einstein sector of the theory, it is clear that
the terms containing derivatives of the curvatures in (\ref{prepoDavid})-(\ref{RenormalizedCharge}) vanish identically. In a similar way, it is easy to
notice that the mass is coming only from the Gauss-Bonnet contribution to
the charge, what is proportional to the Riemann tensor. This is remarkable, as it manifestly shows that, in this framework, the topological term gathers the relevant physical information. The Ricci
tensor and Ricci scalar terms contribute just with numbers to the squared
bracket in the relation above.

For the timelike Killing vector $\xi =\partial_t$, the charge (\ref{prepoDavid})-(\ref{RenormalizedCharge}) gives the mass of Schwarzschild-AdS black holes,
\begin{equation}
{Q_{\rm{top}}^0[\partial_t ]}=m\left[ 1+2\Lambda (\alpha +4\beta )\right] \,, \label{Lomismo}
\end{equation}%
which is in exact agreement with the result (\ref{Mass}) obtained by the ADT method.
A similar result holds for the angular momentum of Kerr-AdS black holes.

The fact that we are able to recover the same expression (\ref{Mass}) from a procedure that is
intrinsically non-perturbative may shed light on the general problem of linear instability of QCG at critical
points. In particular, in the case of Critical Gravity, the above
formula implies the exact vanishing of all conserved quantities for any Einstein space \cite{Anastasiou:Olea:Rivera}.

The contributions of the higher-curvature terms, including the topological term (\ref{GaussBonnet}), also produce a change in the Bekenstein-Hawking entropy of the black holes. Computed as a charge at the horizon, the entropy gets modified in two ways: {apart} from the modification of the prefactor $1/(4G)$ in the area law, which in the presence of $R^2$ terms gets multiplied by a factor $[1+2\Lambda (\alpha +4\beta )]$ in such a way that the first principle of the black hole mechanics still holds, the presence of the topological term induces an extra shift of the entropy formula, which receives an extra constant piece $4\pi\gamma/G$. The latter, being a constant, does not affect the first principle.

\subsection{Gravitational waves}

Now, let us {focus to} other solutions of the theory. Consider non-Einstein spaces that solve the EOM and represent gravitational waves. These are given by a metric of the form \cite{Podolsky}
\begin{equation}\label{Siklos}
ds^2=\frac{\ell^2}{z^2}\left[-(1+F(t,z,x))dt^2+2dtdu+dz^2+d{x}^2\right]\,,
\end{equation}
where $F$ is a function that does not depend on the lightlike coordinate $u$. The coordinates can be taken as $u\in \mathbb{R}$, $t\in \mathbb{R}$, $z\in \mathbb{R}_{> 0}$, and $x\in \mathbb{R}$. Function $F$ can be thought of as describing the profile of the gravitational wave. $F=const$ corresponds to AdS$_4$ spacetime in Poincar\'e coordinates, with the boundary being located at $z=0$. Actually, $z$ relates to the standard radial coordinate of AdS$_4$ by $z=\ell^2/r$.

For {the} metric of the form (\ref{Siklos}), the field equations reduce to
\begin{equation}\label{SiklosEOM}
(\Box-\tilde{m}^2)\Box F=0\,,
\end{equation}
with the effective mass parameter
\begin{equation}\label{squaredM}
\tilde{m}^2=\frac{6(\alpha + 4\beta ) -\ell^2}{\alpha}\,.
\end{equation}

A detailed analysis of geometries (\ref{SiklosEOM}) is given in Ref.\cite{Gaston}, where two sectors of solutions are seen to appear. The functions that solve the second order equation $\Box F =0$ in (\ref{SiklosEOM}) correspond to the cases $F(z)=z^3$ and $F=const$. These are solutions of Einstein type and describe the propagation of a massless mode on AdS$_4$. In addition, the theory contains non-Einstein solutions of the form $F(z)=z^k$ with
\begin{equation}\label{kvalue}
k_{\pm }=\frac32\pm\sqrt{\frac94+\frac{6(\alpha + 4\beta )-\ell^2}{\alpha}}\ ,
\end{equation}
which are actually solutions to $(\Box -\tilde{m}^2)F=0$. The latter correspond to exact gravitational waves solutions that propagate on AdS$_4$ as if they {have} an inertial mass given by (\ref{squaredM}). The above values apply to the propagation of massive AdS waves. Because $k_+>k_-$, only solutions of the form $F(z)=z^{k_+}$ can be asymptotically, locally AdS$_4$. The condition for this to happen is $8(\alpha+3\beta)\geq \ell^2$. If this is obeyed, then a natural question is whether the gravitational energy associated to these stationary waves on AdS$_4$ is actually finite. Remarkably enough, using ADT method one obtains that the energy associated to such solutions, considering the configuration $F=0$ as the background, yields a vanishing result, ${Q_{\rm{ADT}}^0}[\partial_{t}]=0$. This may seem puzzling as the value $\tilde{m}^2\neq 0$ endows the wave solution with inertial effective mass. Still, the net gravitational ADT energy is zero. This phenomenon is analogous to what has been observed with evanescent gravitons in Warped-AdS spaces. In order to check the consistency of our method with the ADT formalism, we should be able to show that (\ref{prepoDavid})-(\ref{RenormalizedCharge}) also yields a vanishing result.

The computation of the total energy of the massive AdS wave solution $F(z)=z^{k_+}$ using (\ref{prepoDavid})-(\ref{RenormalizedCharge}) yields the following result for the prepotential,
\begin{equation}\label{totprep}
q^{\alpha\beta}_{_{\text{top}}}={z^{k+1}}{\ell^{-4}}k(\alpha k^2-3\alpha k+\ell^2-6\alpha-24\beta)\ \delta^{[\alpha\beta]}_{[uz]}\,,
\end{equation}
which precisely vanishes when $k=k_{\pm}$. Therefore, the gravitational energy associated to the solutions $F(z)=z^{k_+}$ computed with (\ref{prepoDavid})-(\ref{RenormalizedCharge}) turns out to be zero as well, namely ${ Q^0_{\rm{top}}}[\partial_{t} ]=0$. This shows the agreement with the ADT method in a case of a non-Einstein solution of the QCG theory.

\subsection{Logarithmic modes}

One of the special features that the higher-curvature theories exhibit at the critical points is the emergence of low-decaying modes. These are described by asymptotically AdS solutions with logarithmic falling-off in the radial direction. Such modes appear, for example, in the chiral point of Topologically Massive Gravity \cite{Log4} and they were also observed in Critical Gravity in dimension four and higher \cite{Log3}. Logarithmic modes in asymptotically AdS space are  a symptom of non-unitarity in the dual CFT.

There exist exact solutions to higher-curvature gravity exhibiting logarithmic decay \cite{Log1,Log2}. These are given by gravitational waves of the type (\ref{Siklos}) with a profile function $F(r)\propto \log r$. Such modes are present, for example, when the coupling constants in the action obey the special relation
\begin{equation}
 6(\alpha+4\beta)=\ell^2 , \label{iopi}
\end{equation}
with $\Lambda=-3/\ell^2$. In particular, this includes Critical Gravity \cite{Lu:Pope}, whose coupling constants satisfy $\alpha =-3\beta =3/(2\Lambda )$; see \cite{Log5} for a general description. When Eq.(\ref{iopi}) holds, the profile function $F(r)$ of the metrics (\ref{Siklos}) take the form $F(r)=c_0 + c_1 \log r$, where $c_0$ and $c_1$ are two arbitrary constants. These modes, when $c_1\neq 0$, correspond to non-Einstein spaces that solve the higher-curvature equations of motion with negative cosmological constant. Since these logarithmic modes decay slowly in asymptotically AdS spaces, then it is natural to ask whether this makes the energy formula to diverge. Remarkably enough, a direct evaluation of the expression $q^{\alpha\beta}_{_{\text{top}}}$ for the logarithmic solution shows that the prepotential identically vanishes by virtue of Eq.(\ref{iopi}). Therefore, as it happens with the gravitational waves discussed in the previous subsection, the computation of gravitational energy associated to the logarithmic solutions is actually zero, despite the weakened asymptotics.

\subsection{Other asymptotics}

There exist other non-Einstein solutions to QCG which describe black holes with different asymptotics. At large distances, such solutions tend to spaces that are not locally AdS$_4$ but have fewer isometries. An example whose metric is known analytic is the Lifhsitz black hole solution \cite{Cai}
\begin{equation}
ds^2= - \frac{r^{2z}}{\ell^{2z}} \left( 1-\frac{r_H^n}{r^n}\right) dt^2 + \frac{\ell^2}{r^2} \left( 1-\frac{r_H^n}{r^n}\right)^{-1} dr^2 + \frac{r^2}{\ell^2}\, (dx^2 + dy^2)\, ,\label{Cai}
\end{equation}
where $r_H$ is an integration constant that gives the location of the horizon and $\ell^2$ is the curvature radius fixed in terms of the cosmological constant $\Lambda$ (see below). $z$ is the so-called dynamical exponent (typically $z\geq 1$), and $n$ is a real number (we assume $n\geq 2$). Coordinates are taken to be $t\in \mathbb{R}$, $r\in \mathbb{R}_{\geq 0}$, $x\in \mathbb{R}$, $y\in \mathbb{R}$.

Metric (\ref{Cai}) plays an important role in the holographic description of non-relativistic condensed matter systems. Asymptotically, (\ref{Cai}) enjoys anisotropic scale invariance under the transformation $t\to \lambda^{z}t$, $r\to \lambda^{-1}r$, $x\to \lambda x$, $y\to \lambda y$, which comes to realize geometrically the symmetry that the so-called fixed Lifshitz points exhibit \cite{Kachru}. The presence of a horizon at $r=r_H$ introduces finite temperature in the problem.

Remarkably, metric (\ref{Cai}) is a solution to the QCG theory \cite{Cai} provided $\alpha=0$ and $\beta = -1/(8\Lambda)=\ell^2/33$, and for the specific value $z=3/2$ of the dynamical exponent and $n=3$.

In order to use our method to compute the gravitational energy associated to (\ref{Cai}), we need to set the value of the couplings constant of the topological term, $\gamma$. As for the case of asymptotically AdS$_4$ spaces, we can fix the value of $\gamma $ by requiring the variation of the action on the background geometry to vanish. In the case of Lifshitz, the background geometry corresponds to the solution with $r_H=0$. The latter geometry is singular as it contains incomplete timelike geodesics. However, all the curvature invariants associated to it turn out to be constant, and in particular $R=-33/(2\ell^2)$. This yields {the first term $(\frac{1}{2}+\beta R)\delta _{\alpha \beta }^{\mu \nu }$ in $E_{\alpha \beta }^{\mu \nu }$ to vanish, giving rise to $E_{\alpha \beta }^{\mu \nu }=\frac{\gamma }{2}\,\delta _{\alpha \beta \gamma \delta }^{\mu \nu \tau \sigma}R_{\tau \sigma }^{\gamma \delta }$ proportional to $\gamma $, whose divergence vanishes, $\nabla ^{\mu }E_{\mu \nu }^{\alpha \beta }=0$.} Then, the surface term $\Theta^{\alpha}_{_{\text{total}}}$ simply reduces to
\begin{equation}
\Theta^{\alpha}_{_{\rm{total}}}=\gamma g^{\nu \lambda }\delta\Gamma_{\beta \lambda}^{\mu}\delta_{\mu\nu\sigma\lambda}^{\alpha\beta\gamma\delta}R_{\gamma\delta}^{\sigma\lambda},
\end{equation}
which implies $\gamma =0$. Therefore, the prepotential reduces to the one of Iyer-Wald, $q_{_{\rm{top}}}^{\alpha \beta}=q_{_{\rm{IW}}}^{\alpha \beta}=0$, and thus the mass of the Lifshitz black hole (\ref{Cai}) is found to be zero, ${Q_{_{\rm{top}}}^0}[\partial_t]=0$, in agreement with the result of \cite{Cai}.

The fact that the prepotential vanishes expresses that the solution of \cite{AGGH} exists at a singular point in the parameter space where the action identically vanishes. It is then natural to ask what is the result for the case of non-Einstein black hole solutions like (\ref{Cai}) but for which the action does contribute. It was shown in \cite{AGGH} that QCG also admits solutions of the type (\ref{Cai}) at a different point of the parameter space. More precisely, a black hole solution with $z=6$ and $n=4$ exists provided $\alpha=-(5/3)^2\beta=(5\ell/8)^2$ and $\Lambda = -51/(2\ell^2)$, and this solution, unlike the one of \cite{Cai}, has non-zero action. It can be shown that for such a solution the topological term (\ref{GaussBonnet}) does not suffice to regularize the action and provide a gravitational energy computation. This manifestly shows how non-trivial is the fact that such a method works well for asymptotically AdS spaces. An underlying reason lies in conformal properties of asymptotically Einstein spaces, for which the contribution of the topological term turns out to agree with those entering in the the so-called Ashtekar-Magnon-Das conformal mass \cite{AMD1,AMD2}. We will not discuss the precise relation with this here, see Refs.\cite{YPang,Araneda}.

\section{Conclusions}

Motivated by the problem of linear instabilities of higher-curvature gravity and by the importance such corrections to Einstein theory have in the context of quantum gravity, in this paper we gave a definition of gravitational energy for such a theory including arbitrary {quadratic-curvature} corrections to Einstein equations. We focused on the case of four-dimensional theories in presence of negative cosmological constant, and with asymptotic AdS boundary conditions. For these theories, we computed the gravitational energy and angular momentum of black holes, obtaining results in perfect agreement with previous computations performed using different methods. Our method, however, is intrinsically non-linear as it relies on the idea of adding to the gravity action topological invariant terms which suffice to regularize the Noether charges and render the variational problem well-posed. This makes our method well-suited to study the higher-curvature theories at the so-called critical points, where linear instabilities arise. The idea of adding to the gravitational action topological invariants to regularize the Noether charges had been previously developed by one of the authors and collaborators in the case of second order theories, such as general relativity, and here we have shown how this can be extended to higher-derivative theories. In addition to black holes, we discussed other solutions, such as gravitational waves in AdS. This enabled us to probe the consistency of our approach in the non-Einstein sector of the quadratic-curvature theories.

Several open questions remain for future investigations. For instance, it would be interesting to explore the extension of this method to higher-derivative theories in higher (even) dimensions, where the {quadratic-}curvature corrections introduce new qualitative features. It would also be interesting to investigate this method in the context of AdS/CFT holography; in particular, the connection between topological terms and the renormalization of holographic correlation functions calculated for this gravity theory,
along the lines of \cite{Miskovic:Olea,Miskovic:Olea:Tsoukalas,Johansson,Anastasiou:Olea}. On general grounds, the study of higher-curvature gravity in the context of AdS/CFT is important as fourth-order field equations introduce new sources at the boundary. It is then required to fully understand whether the addition of a topological invariant always produces a variational principle which is finite and expressible in terms of sources kept fixed at the conformal boundary.
%In this framework, the addition of topological terms to higher-derivative gravity actions is an important step forward, as it was long believed that topological regularization of AdS gravity was an exclusive feature of theories with second-order field equations.

\acknowledgments

This work was supported in parts by the NSF through grant PHY-1214302, Chilean FONDECYT project N$^{\circ}$1170765 and the grant VRIEA-PUCV N$^{\circ}$039.314/2018. D.R.B. is a UNAB M.Sc. Scholarship holder.

\end{document}